\documentstyle[emulateapj,onecolfloat,psfig]{article}
\newcommand{\bn}{\hat{\bf n}}
\newcommand{\bl}{{\bf l}}
\newcommand{\bll}{{\bf L}}
\newcommand{\tot}{{\rm tot}}

\newcommand{\ApJ}{ApJ}

\newcommand{\PRD}{Phys. Rev. D}

\newcommand{\etal}{et al.}

\newcommand{\AsAs}{A\&A}
\newcommand{\amp}{\& }
\newcommand{\aut}[2]{{#1, #2.}}
\newcommand{\laut}[2]{{#1, #2.}}
\newcommand{\refs}[6]{#5, #2, #3,  {#4}.}

\newcommand{\mybib}[2]{\bibitem[#1]{#2}}

\begin{document}
\twocolumn[
\title{Mapping the Dark Matter through the CMB Damping Tail}
\author{Wayne Hu}
\affil{Department of Astronomy and Astrophysics, University of Chicago,
Chicago, IL 60637\\
}

\begin{abstract}
The lensing of CMB photons by intervening large-scale structure leaves
a characteristic imprint on its arcminute-scale anisotropy that can be
used to map the dark matter distribution in projection on degree scales
or $\sim 100 h^{-1}$ Mpc comoving. We introduce a new algorithm for
mass reconstruction which optimally utilizes information
from the weak lensing of CMB anisotropies in the damping tail.  
Individual degree-scale mass structures can be recovered with
high signal-to-noise from a foreground-free 
CMB map of arcminute scale resolution, 
specifically with a FWHM beam $< 5'$ and a noise level
$< 15$ ($10^{-6}$-arcmin) or 41 ($\mu$K-arcmin).   
\end{abstract}

]

\section{Introduction}

It is well-known from the study of the weak gravitational lensing of faint
galaxies that the distortion of background images can be used to map
the intervening mass distribution in projection (\cite{TysWenVal90} 1990;
\cite{KaiSqu93} 1993).  As the most distant background image available,
maps of the CMB temperature distribution provide a unique opportunity to
map the distribution of dark matter.
They provide information about structures on the largest linear scales 
in the high redshift universe (\cite{ZalSel99} 1999)  
and hence complement information from weak
lensing surveys. 
The main difficulty is that unlike an image of background galaxies, 
the temperature distribution of the CMB is to good approximation a 
Gaussian random field with no characteristic shape.

Algorithms in the literature for extracting the intervening mass 
distribution from lensed CMB maps have shown the potential for
statistical detections by the 
Planck 
satellite.\footnote{http://astro.estec.esa.nl/SA-general/Projects/Planck}
\cite{Ber98} (1998) considered the distortion to the Hessian of
the temperature field. \cite{ZalSel99} (1999) considered distortions to
the product of gradients of the temperature field.  In neither case is
it possible to extract high signal-to-noise maps of the dark matter.

Recently \cite{Zal00} (2000) showed that the damping tail of 
CMB anisotropies (see e.g. \cite{HuWhi97a} 1997) exhibits enhanced
lensing effects in the four-point function.
Indeed even the two-point function or power spectrum shows enhanced
effects in this region due to the multitude of acoustic peaks and the
sharp decline in intrinsic power associated with damping
(\cite{MetSil97} 1997).  

\cite{Hu01} (2001) showed that there is
a quadratic estimator that recovers all of the information in the four-point
function about the mass distribution on large scales.  Even for
experiments like Planck that only partially resolve the damping tail,
this estimator reduces the noise variance
of the recovered projected mass power spectrum by over an order of magnitude.
Planned experiments to measure arcminute-scale secondary CMB anisotropies 
can potentially 
use this statistic to map the dark matter at high signal-to-noise.

We begin in \S \ref{sec:lensing} by reviewing the effect of lensing on
CMB temperature maps.  In \S \ref{sec:reconstruction}, we describe
the reconstruction algorithm and test it with realizations of the CMB
temperature field and instrumental noise. We discuss observational
strategies for optimizing the reconstruction in \S \ref{sec:discussion}.
For illustration purposes we use a flat $\Lambda$CDM cosmology throughout with
parameters 
$\Omega_c = 0.3$, $\Omega_b=0.05$, $\Omega_\Lambda=0.65$, $h=0.65$, $n=1$
and $\delta_H=4.2\times 10^{-5}$.

\section{Lensing}
\label{sec:lensing}

\begin{figure}[t]
\centerline{\epsfxsize=3.5truein\epsffile{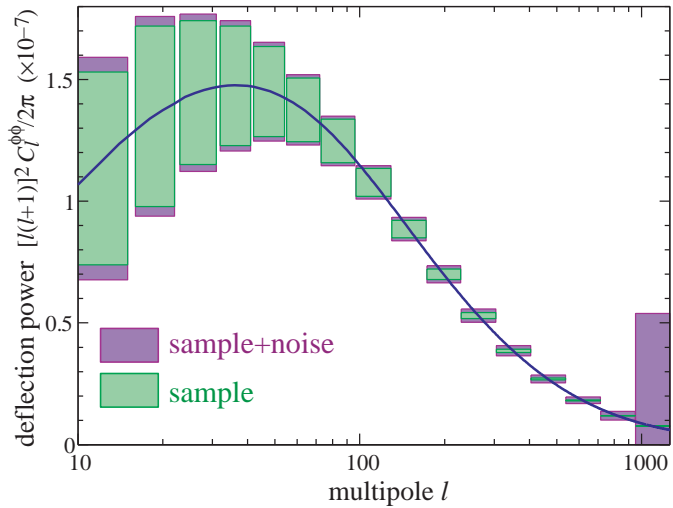}}
\caption{\footnotesize Lensing deflection power spectrum for the $\Lambda$CDM model.  Error
bars represent the total (sample plus noise) variance and sample variance
from recovery from an area of $f_{\rm sky}=0.1$ and an experiment with
a beam of $\sigma=1.5'$ and noise $w^{-1/2}=10$ $(10^{-6}$-arcmin).  Note that
the errors for $L \lesssim 200$ are dominated by sample variance implying that the recovered map
has high signal-to-noise.}
\label{fig:defl}
\end{figure}

Weak lensing of the CMB photons by the intervening mass distribution
remaps the primary temperature field $\tilde \Theta(\bn)$ as a function of
the directional vector $\bn$ on the sky as (e.g. \cite{Sel96} 1996; 
\cite{GolSpe99} 1999)
\begin{equation}
\Theta(\bn) = \tilde \Theta(\bn + \nabla \phi)\,,
\label{eqn:remapping}
\end{equation}
where $\nabla \phi$ is the deflection angle which is related
to the gravitational potential $\Psi({\bf x},D)$ as
\begin{equation}
\phi(\bn) = -2 \int d D\, {D_A(D_s-D) \over D_A(D)\, D_A(D_s)} \Psi(D \bn,D)\,,
\end{equation}
where $D$ is the comoving coordinate distance along the line of sight and $D_A$
is the comoving angular diameter distance associated with $D$. $D_s$ is
the coordinate distance to the last scattering surface.  
In a flat universe $D_A=D$.   The projected potential $\phi(\bn)$ has a power spectrum $C_L^{\phi\phi}$ which is itself a projection of the gravitational
potential power spectrum (see e.g. \cite{Hu00b} 2000, Eqn. 28).  
For the scales of interest here, the gravitational potential power spectrum
is in the linear regime and hence $\phi(\bn)$ is a Gaussian random field.
The power spectrum
of the deflection angles is given by $L(L+1) C_L^{\phi\phi}$ (see
Fig.~\ref{fig:defl} for $\Lambda$CDM).  
For reference, the convergence power spectrum, or projected mass, is 
given by $[L(L+1)]^2 C_L^{\phi\phi}/4$.  The rms deflection angle
\begin{equation}
\theta_{\rm rms}^2 = \sum_{L=1}^{\infty} {2L+1 \over 4\pi} L(L+1)C_L^{\phi\phi}
\end{equation}
is $2.6'$ for this model but its coherence scale corresponds to the peak of
the log power spectrum at $L \sim 60$ or a few degrees.  Counterintuitively then,
the arcminute scale structure of the CMB temperature field yields information 
about the mass distribution on much larger linear scales.  
The deflection power comes mainly
from structures at redshifts $z \sim 1-2$ and scales of  
$k \sim$ few $\times 10^{-2}$ Mpc$^{-1}$ (see \cite{ZalSel99} 1999, Fig.~7). 

\begin{figure}[t]
\centerline{\epsfxsize=2.75truein\epsffile{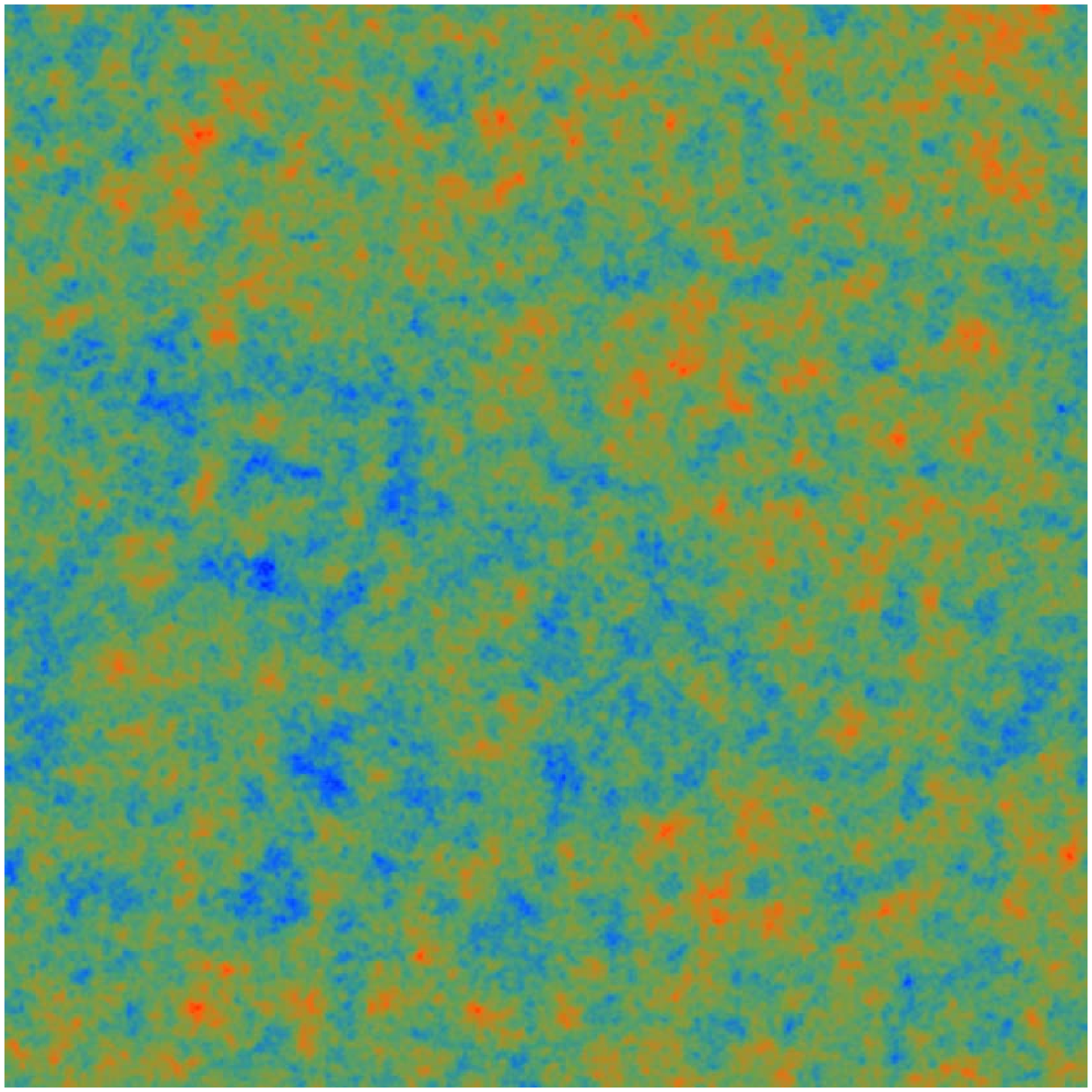}}
\vskip 0.1cm
\centerline{\epsfxsize=2.75truein\epsffile{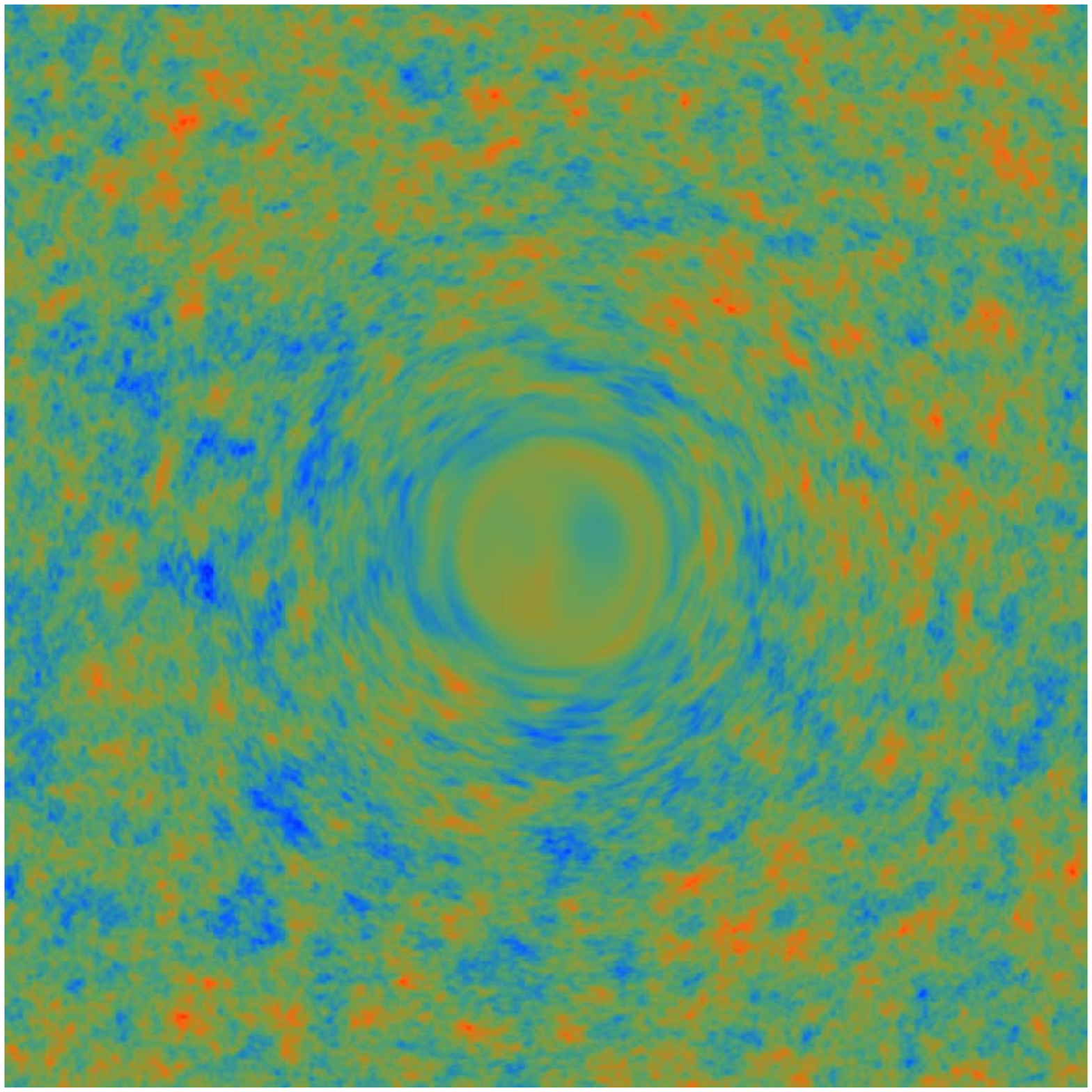}}
\caption{\footnotesize Top: A $32^\circ \times 32^\circ$ realization of the 
CMB temperature field. Bottom: A toy example of the lensing effect. A
circularly symmetric projected mass with deflection angles comparable to
the size of the structure.  The distortion of the fine-scale anisotropy of
the CMB traces the lensing structure on much larger scales.}
\label{fig:lensing}
\end{figure}

To simulate a lensed CMB map, one makes
a Gaussian random realization of the unlensed CMB power spectrum $\tilde C_l$ and remaps 
the temperature field according to a random realization of the projected potential
$C_L^{\phi\phi}$.  Detector noise and residual foregrounds are then added as
a realization of $C_l^{\rm noise}$.  For detector noise
and a finite beam of $\sigma$ (FWHM) (\cite{Kno95} 1995)
\begin{equation}
C_l^{\rm noise} = w^{-1} e^{l(l+1)\sigma^2 /8\ln 2}\,,
\end{equation}
with $w^{-1}$ is the noise in units of ($\Delta T/T$-radian)$^2$.  

Because the deflection angles are small compared with the scale of
the structures, the lensing effect is difficult to see directly in a map.  To gain a better
intuition for the nature of the effect, let us first consider lensing by a circularly
symmetric Gaussian profile in projected mass with a scale of 
$5^\circ$ and an amplitude corresponding to 
a $3^\circ$ maximum deflection. 
As in the case of the weak lensing of faint galaxies,
the distortion represents a tangential shearing of the image.  Unlike the faint galaxy case,
the source image is a Gaussian random field.  
Although the temperature map itself
clearly shows evidence for lensing, the two point statistics of
the CMB temperature field do not suffice to reconstruct the mass distribution of the lens. 

\section{Reconstruction}
\label{sec:reconstruction}

The case of the symmetric lens in Fig.~\ref{fig:lensing} suggests that a statistic
related to the Laplacian of the temperature field would trace the underlying
mass distribution.  The fact that both hot and cold spots are lensed alike
washes out the signal in the Laplacian itself.  \cite{Hu01} (2001) showed 
that a related statistic, the divergence of the temperature-weighted 
gradient of the map retains all of the information inherent in the four-point
function (\cite{Zal00} 2000).  
Here we consider its use in mapping the dark matter.

We describe the technique for reconstructing the dark matter field on
small sections of the sky $\theta_{\rm map}<60^\circ$ for which spherical
harmonic analysis can be replaced by Fourier techniques.  For the
generalization to the curved sky see \cite{Hu01} (2001). 

The first step is to take the gradient of the 
temperature map by filtering in the Fourier domain,
\begin{equation}
{\bf G}(\bn) = \int{d^2 l \over (2\pi)^2}\, i\bl {\tilde C_l \over C_l^\tot} 
		\Theta(\bl)e^{i \bl \cdot\bn} \,,
\end{equation}
where $C_l^\tot = C_l + C_l^{\rm noise}$, $C_l$ ($\tilde C_l$) is lensed 
(unlensed) power spectrum. 
Note that taking the gradient effectively high-pass
filters the map.
Next construct an explicitly high-pass filtered temperature map
\begin{equation}
W(\bn) = \int{d^2 l \over (2\pi)^2}\, {1 \over C_l^\tot} 
		\Theta(\bl)e^{i \bl \cdot\bn} \,,
\end{equation}
to weight the gradient
\begin{equation}
\tilde{\bf G}(\bn) = W(\bn) {\bf G}(\bn)\,.
\end{equation}
Finally take a filtered divergence of this field in the Fourier domain
\begin{equation}
D(\bn) = -\int{d^2 L \over (2\pi)^2}\, 
		{N_L \over L}
		i\bll \cdot\tilde {\bf G}(\bll)e^{i \bll \cdot\bn} \,.
\end{equation}

\begin{figure}[t]
\centerline{\epsfxsize=2.75truein\epsffile{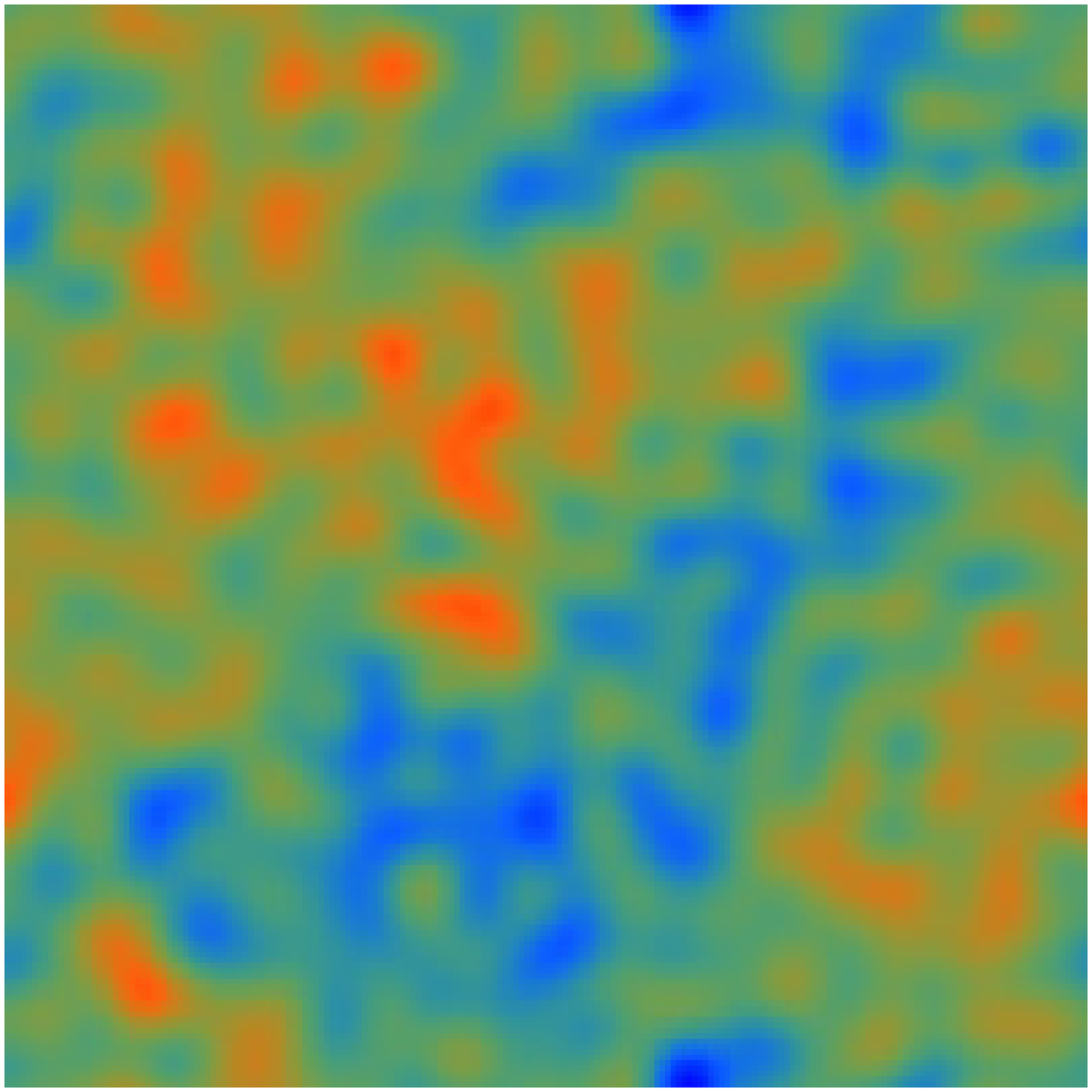}}
\vskip 0.1cm
\centerline{\epsfxsize=2.75truein\epsffile{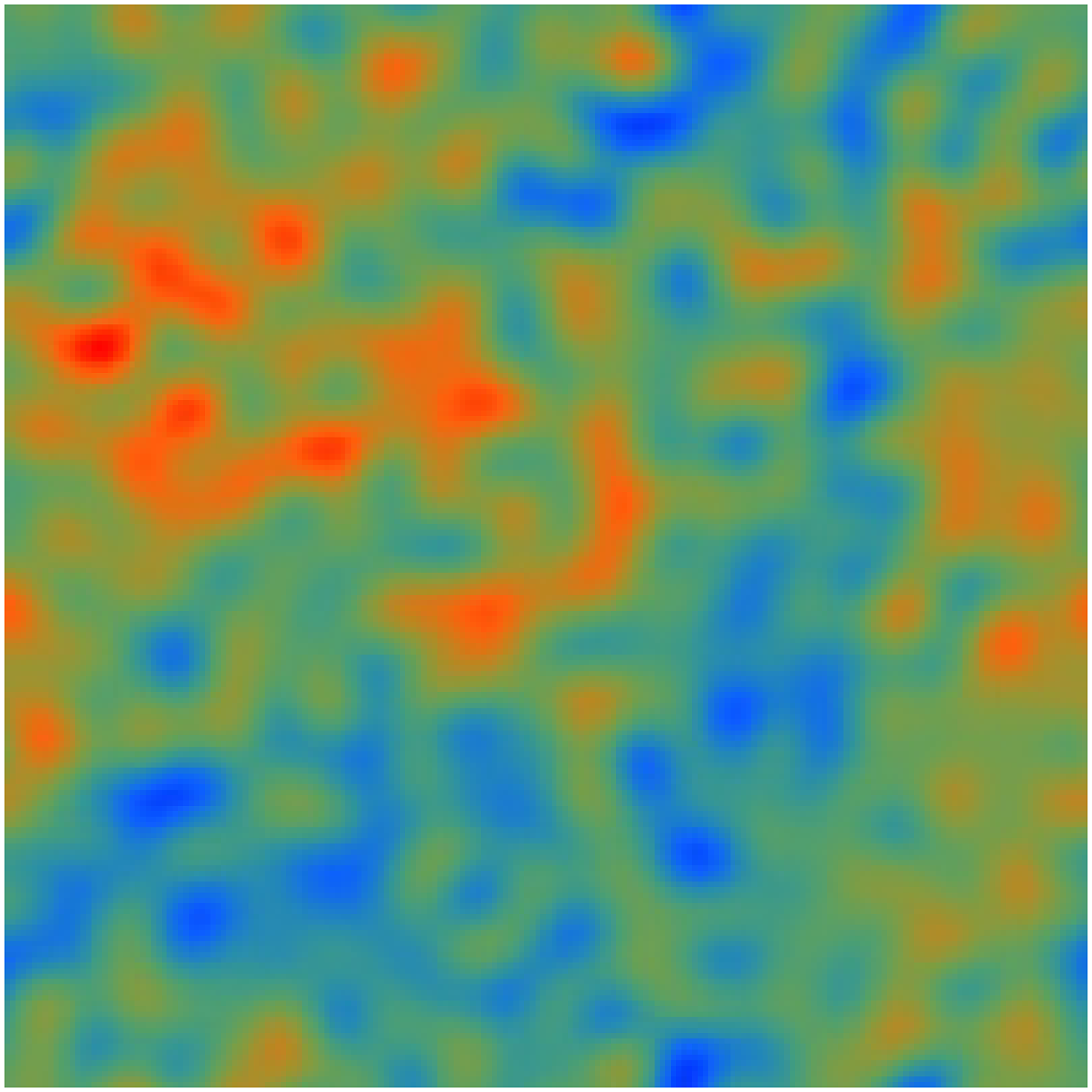}}
\caption{\footnotesize Top: A $32^\circ \times 32^\circ$ realization of the deflection 
field in the $\Lambda$CDM model.  Bottom: Recovery of the deflection field with a $\sigma=1.5'$ beam
(FWHM) and detector noise of $w^{-1/2} = 10$ $(10^{-6}$-arcmin).}
\label{fig:reconstruct}
\end{figure}

The normalization factor $N_L/L$ may be chosen so that $D(\bn)$ averaged
over an ensemble of CMB realizations recovers the deflection field
\begin{equation}
d(\bn) \equiv \int{d^2 L \over (2\pi)^2}\, L \phi(\bll) e^{i \bll \cdot\bn} \,.
\end{equation}
To determine $N_L$ consider the operations directly in the Fourier domain,
\begin{equation}
D(\bll) =  {N_L \over L} \int {d^2 l_1 \over (2\pi)^2}
		(\bll \cdot \bl_1\, \tilde C_{l_1} +  \bll \cdot \bl_2\, 
		\tilde C_{l_2})
		{\Theta_{l_1} \Theta_{l_2} \over 
		2 C_{l_1}^\tot C_{l_2}^\tot}\,,
\end{equation}
where $\bl_2=\bll-\bl_1$.  Taylor expanding Eqn.~(\ref{eqn:remapping}) for
the lensing one obtains (\cite{Hu00b} 2000)
\begin{equation}
\Theta(\bl) = \tilde \Theta(\bl) - \int {d^2 l_1 \over (2\pi)^2}
	\Theta(\bl_1) \phi(\bl-\bl_1) (\bl - \bl_1) \cdot \bl_1\,,
\end{equation}
so that
\begin{eqnarray}
\left< D(\bll) \right>_{\rm CMB} = d(\bll)=  L \phi(\bll)\,,
\end{eqnarray}
if
\begin{equation}
N_L^{-1} = {1 \over L^2} \int {d^2 l_1 \over (2\pi)^2}
		{(\bll \cdot \bl_1\, \tilde C_{l_1} +  \bll \cdot \bl_2\, 
		\tilde C_{l_2})^2 \over 2 C_{l_1}^\tot C_{l_2}^\tot}\,.
\end{equation}
Notice that the filters are designed so that the lensing effects in
the Fourier domain add coherently such that the dot product above
comes in as the square.  This is a reflection of the optimization.

We show an example of this reconstruction on a $32^\circ \times 32^\circ$ field
in Fig.~(\ref{fig:reconstruct}) with
detector noise added as appropriate 
for a beam of $\sigma=1.5'$ and noise of $w^{-1/2}=
10$ $(10^{-6}$-arcmin) or (27$\mu$K-arcmin) 
additionally low pass filtered to show $L\le 150$ where the
signal-to-noise is the highest.  Alternately, Weiner filtering can be used
to get a better visual impression of the fidelity of the map.
In any case, the degree scale features in the map are recovered at 
good signal-to-noise.

In the idealization of an ensemble of CMB maps lensed
by the {\it same} structure, $D(\bll)$ returns an unbiased estimate of the 
deflection
map.  However given that we only have one realization of the lensing per
lens, it is important to understand the properties
of the noise introduced by the Gaussian primary anisotropies themselves and
the instrumental and/or foreground noise.  Following \cite{Hu01} (2001),
\begin{equation}
\langle D^*(\bll) D(\bll')\rangle = (2\pi)^2 \delta(\bll-\bll') 
	\left( L^2 C_L^{\phi\phi} + N_L \right)\,,
\end{equation}
so that $N_L$ also plays the role of the noise power spectrum.  A deflection
power spectrum extracted from this statistic must remove this noise bias.
As discussed in \cite{Hu01} (2001), the noise bias may alternately 
be eliminated by cross-correlating
maps reconstructed from independent $l$-bands in the original lensed map.

Under the assumption of Gaussian statistics, the signal-to-noise
per $L$ in the deflection power spectrum is given by 
\begin{equation}
\left( {S \over N} \right)_L^2 = f_{\rm sky} {2 L+1 \over 2}
\left( L^2 C_L^{\phi\phi} \over {L^2 C_L^{\phi\phi} + N_L} \right)^2	\,,
\label{eqn:sn}
\end{equation}
and the precision with which the binned deflection power spectrum 
can be recovered by an experiment with a sky fraction of $f_{\rm sky}=0.1$ 
($\sim 4000$deg$^2$) and $\sigma=1.5'$, $w^{1/2}=10$ $(10^{-6}$-arcmin) is shown
in Fig.~\ref{fig:defl}.  Also shown are the errors provided by 
sample variance alone.  Since sampling errors dominate 
at $L \lesssim 150$, the recovered map
has good signal-to-noise on those characteristic structures.
This is independent of the actual sky fraction covered by the experiment.

\section{Discussion}
\label{sec:discussion}

The statistic introduced here utilizes CMB structures in the 
{\it arcminute} regime of the damping tail to map the 
dark matter on {\it degree} scales.  
As with the weak lensing of
faint galaxies, image distortions manifest on small angular 
scales are used to reconstruct the mass on a much larger scale.
Mapping the dark matter distribution therefore requires
high resolution, high signal-to-noise maps of the CMB anisotropies themselves.
Conversely, though a wide field of at least several degrees 
on the side is required
to map the full extent of the structures expected, 
the statistic essentially high
pass filters the input CMB maps.  A true map that retains correlations across
these scales is not necessary.

\begin{figure}[t]
\centerline{\epsfxsize=3.5truein\epsffile{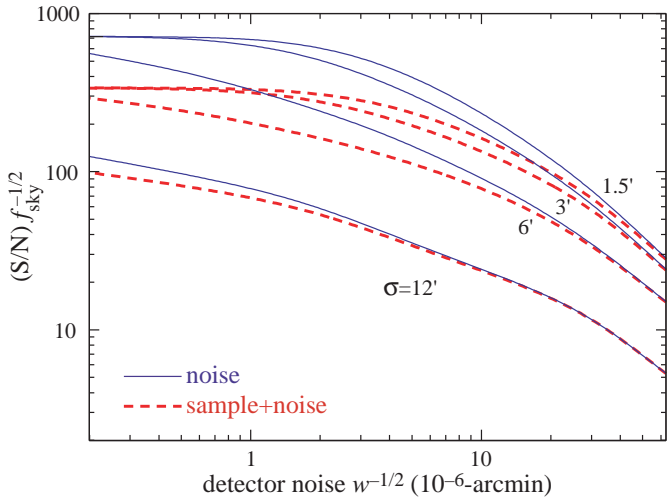}}
\caption{\footnotesize Signal-to-noise as a function of detector noise, beam and sky
fraction $f_{\rm sky}$.  Solid lines include both sample and Gaussian 
noise (primary CMB and instrumental) variance; dashed lines include
only Gaussian noise variance.  The signal-to-noise drops off 
rapidly for $w^{-1/2}>15$ $(10^{-6}$-arcmin) and FWHM beams $\sigma>5'$.}
\label{fig:sensitivity}
\end{figure}

To see how an observing strategy might be optimized for mapping the dark 
matter, let us consider the trade-offs between sky coverage, instrumental noise and
beam.  Because this statistic is a quadratic function of the temperature
fluctuation data, the balance differs from the usual case. 
In Fig.~\ref{fig:sensitivity}, we show the total signal-to-noise
in the measurement of the deflection power spectrum 
(summed in quadrature over $L$) of an experiment 
as a function of these parameters.  We consider separately the case of 
noise variance
from the Gaussian random primary anisotropies and detector noise alone and
combined with the sample variance of the lensing fields.  When the
former exceeds the latter, a high signal-to-noise map of the structures
results.  Because this is an integrated statistic,
the characteristic signal-to-noise for large-scale features is much higher
(see Fig.~\ref{fig:defl}).

Compare the steep increase in the signal-to-noise as the detector noise
is reduced with the shallow increase with sky coverage of
$f_{\rm sky}^{1/2}$.  Up until $w^{-1/2} \sim 10$ $(10^{-6}$-arcmin), observing
time is best spent going deep rather than wide.  Beyond this point, the
intrinsic noise variance provided by the primary CMB anisotropies themselves
begins to dominate and saturate the signal-to-noise.  If the goal is
to produce a high signal-to-noise map of structures, then going down to
$w^{-1/2} \sim 1$ $(10^{-6}$-arcmin) can achieve substantially improved maps of the finer scale
structures in the map.  Another crucial factor is the
beam size.  To resolve the structures that best trace the lensing, a beam
of $\sigma < 5'$ is required and it is not until $\sigma \sim 1'-2'$ that
the gains saturate.  If foregrounds 
are not removed from the map through their
spatial coherence and/or frequency dependence then this balance can shift to 
larger angular scales and more sky coverage.  For the $1.5'$, $10$ $(10^{-6}$-arcmin)
baseline experiment, inclusion of Gaussian random noise from the 
Sunyaev-Zel'dovich and
Vishniac effects in $C_l^{\rm tot}$ imply a relative degradation in 
signal-to-noise of $\sim 10\%$ and $\sim 1\%$ (for $\tau=0.1$) 
respectively and so
do not require substantial reoptimization.

A high signal-to-noise map of the dark matter in projection can also be used
to pull out tracers of the large-scale structure of the universe
in other maps through cross-correlation.  Examples include 
secondary anisotropies such as the integrated Sachs-Wolfe and Sun\-yaev-Zel'dovich effects
(\cite{GolSpe99} 1999; \cite{SelZal98} 1998; \cite{CooHu00a} 2000).  
One can show that the statistic 
employed here retains all of the information in the full bispectrum of 
the secondary-lensing-primary correlation and so is the optimal statistic 
to measure these correlations.

The filters used to reconstruct the dark matter map formally require as input the
power spectrum of the CMB {\it before} lensing.  The lensed CMB power spectrum
will of course be measured to exquisite precision by CMB satellites and by
the input temperature map themselves. 
Employing the lensed CMB power spectrum in the filter or an otherwise
slightly incorrect assumption simply degrades the signal-to-noise by 
a correspondingly small amount but does not introduce spurious 
structures in the ensemble-averaged recovery.  
They appear as a calibration error for the mass map.
Indeed in the context of a parameterized cosmology the
unlensed CMB power spectrum may itself be reconstructed from the observed
spectrum. 
For a non-uniform survey geometry, with perhaps foreground-contaminated
regions removed, more sophisticated techniques than the Fourier-transform
filtering scheme employed here will have to be developed.
These  complications should not present an insurmountable obstacle
to the goal of mapping the dark matter in projection at intermediate redshifts.
 
{\it Acknowledgements:} I acknowledge useful conversations with
A.R. Cooray and M. Zaldarriaga as well as support from NASA NAG5-10840, DOE OJI 
and an Alfred P. Sloan Foundation Fellowship.

\end{document}